\documentclass[pre%
,twocolumn%
,superscriptaddress%
,floats
,aps%
,amssymb%
]{revtex4}
\usepackage{amsmath}%
\usepackage{graphicx}

\begin{document}
\title{Reduction of Two-Dimensional Dilute Ising Spin Glasses} 
\date{\today}
\author{Stefan Boettcher}  
\email{sboettc@emory.edu}
\affiliation{Physics Department, Emory University, Atlanta, Georgia
30322, USA}  
\author{Alexander K. Hartmann}  
\email{hartmann@theorie.physik.uni-goettingen.de}
\affiliation{Institut f\"ur Theoretische Physik, Universit\"at
  G\"ottingen, 37077 Goettingen, Germany}  

\begin{abstract} 
The recently proposed reduction method is applied to the
Edwards-Anderson model on bond-diluted square lattices. This allows,
in combination with a graph-theoretical matching algorithm, to
calculate numerically exact ground states of large systems.
Low-temperature domain-wall excitations are studied to determine the
stiffness exponent $y_2$. A value of $y_2=-0.281(3)$ is found,
consistent with previous results obtained on undiluted lattices.  This
comparison demonstrates the validity of the reduction method for
bond-diluted spin systems and provides strong support for similar
studies proclaiming accurate results for stiffness exponents in
dimensions $d=3,\ldots,7$.
\hfil\break  PACS number(s): 
05.50.+q
, 64.60.Cn
, 75.10.Nr
, 02.60.Pn
.
\end{abstract} 
\maketitle

\section{Introduction}
\label{intro}
Despite more than two decades of intensive research, many properties
of spin glasses~\cite{reviewSG1,reviewSG2,F+H,reviewSG3}, especially
in finite dimensions, are still not well understood. The most simple
model is the Edwards-Anderson model (EA)~\cite{F+H},
\begin{eqnarray}
H=-\sum_{<i,j>}\,J_{i,j}\,x_i\,x_j,\quad(x_i=\pm1),
\label{Heq}
\end{eqnarray}
with Ising spins $x_i=\pm1$ arranged on a finite-dimensional lattice
with nearest-neighbor bonds $J_{i,j}$, randomly drawn from a
distribution $P(J)$ of zero mean and unit variance.

For two-dimensional Ising spin glasses it is now widely accepted that
no ordered phase for finite temperatures
exists~\cite{RSBDJ,kawashima1997,HY,houdayer2001,AMMP}, while spin
glasses order at low temperatures in higher
dimensions~\cite{kawashima1996,marinari1998,Hy3d,CBM}.  This can be
seen e.g.\ by studying the stiffness exponent $y$ (often labeled
$\theta$) which is in many respects one of the most fundamental
quantities to characterize the low-temperature state of disordered
systems.  This exponent provides an insight into the effect of
low-energy excitations of such a system~\cite{FH,BM1}. A recent study
suggested the importance of this exponent for the scaling corrections
of many observables in the low-temperature regime~\cite{BKM}, and it
is an essential ingredient to understand the true nature of the energy
landscape of finite-dimensional glasses~\cite{KM,PY2,PY}.

To illustrate the meaning of the stiffness exponent, one my consider
an ordinary Ising ferromagnet of size $L^d$ having bonds $J=+1$, which
is well-ordered at $T=0$ for $d>1$, with periodic boundary
conditions. If we make the boundary along one spatial direction
anti-periodic, the system would form an interface of violated bonds
between mis-aligned spins, which would raise the energy of the system
by $\Delta E\sim L^{d-1}$. This ``defect''-energy $\Delta E$ provides
a measure for the energetic cost of growing a domain of overturned
spins, which in a ferromagnet simply scales with the surface of the
domain. In a disordered system, say, a spin glass with an equal mix of
$J=\pm1$ couplings, the interface of such a growing domain can take
advantage of already-frustrated bonds to grow at a reduced or even
decreasing cost. To wit, we measure the probability distribution
$P(\Delta E)$ of the interface energies induced by perturbations at
the boundary of size $L$, for which the typical range $\sigma(\Delta
E)=\sqrt{<\Delta E^2>-<\Delta
E>^2}$ of the defect-energy may scale like
\begin{eqnarray}
\sigma(\Delta E)\sim \,L^y.
\label{yeq}
\end{eqnarray}
From the above consideration it is clear that $y\leq d-1$, and a bound
of $y\leq (d-1)/2$ has been proposed for the Edwards-Anderson model
generally~\cite{FH}. Particular ground states of systems with $y\leq0$
would be unstable or only marginally stable with respect to
spontaneous fluctuations, whether induced thermally or
structurally. These fluctuations could grow at no cost, like in the
case of the one-dimensional ferromagnet where $y=d-1=0$. Such a system
does not manage to attain an ordered state for any finite
temperature. Conversely, a positive sign for $y$ at $T=0$ indicates a
finite-temperature transition into an ordered regime while its value
is a measure of the stability of the ordered state. It is generally
believed that the EA possesses a glassy low-temperature regime,
i.~e. $y>0$, for $d\geq3$
\cite{PY,SY,Kirkpatrick,Hy3d,CBM,BM,BC,CB,MKpaper,Hd4,hukushima2000},
while such a phase is absent, and $y<0$, for $d\leq2$
\cite{HY,AMMP,BM,HBCMY}.  A value of $y=0$ marks the lower critical
dimension.

The stiffness exponent provides a measure of the effect of excitations
on a spin system around ground state configurations, induced by
low-temperature fluctuations. It is computationally convenient to
induce such excitations by perturbing the system of size $n=L^d$ at
one boundary and measuring the response for increasing system size
$L$. The square lattices considered here have one open and one
periodic boundary, and we determine the energy difference $\Delta E$
between the ground states of a given bond-configuration, once with
periodic and once with anti-periodic boundary conditions ({\bf P-AP}
method). Anti-periodic boundary conditions are obtained by reversing
the sign of a strip of bonds along the periodic boundary. Using a
symmetric bond distribution $P(J)$, $\Delta E$ will be also
symmetrically distributed, but with a variance $\sigma$ of these
excitation that may scale with $L$ according to Eq.~(\ref{yeq}).

Until recently, the consensus of the results for $d=3$ ranged from
$y_3\approx0.19$ to
$\approx0.27$~\cite{PY,SY,Kirkpatrick,BC,BM,CB,Hy3d,CBM,MKpaper},
while there was only two results in $d=4$, $y_4=0.64(5)$~\cite{Hd4}
and $y_4=0.82(6)$~\cite{hukushima2000}.  In
Refs.~\cite{Stiff1,Stiff2}, it was proposed to study the EA in
Eq.~(\ref{Heq}) on bond-diluted lattices to obtain more accurate
scaling behavior for low-temperature excitations.  One can remove
iteratively low-connected spins from the lattice and alter the
interactions, i.e.\ {\em reduce} the system, in such a way that the
ground-state energy of the reduced system is the same as the original
system. In this way often much larger lattice sizes $L$ can be
simulated compared to undiluted ones and, in combination with
finite-size scaling, enhanced scaling regimes are achieved. In this
manner, improved or entirely new values for the $T=0$ stiffness
exponents in dimensions $d=3$ to 7 were computed for lattices with
{\it discrete} bonds, $\pm J$, resulting in $y_3=0.24(1)$,
$y_4=0.61(2)$, $y_5=0.88(5)$, $y_6=1.1(1)$, and $y_7=1.24(5)$.

The novelty of the procedure used in Refs.~\cite{Stiff1,Stiff2} makes
it difficult to assess the validity and the accuracy of the approach,
since few data of comparable accuracy exists for the results presented
there. The scaling Ansatz used is based on various reasonable but
untested assumptions, for instance that the stiffness exponent does
not depend on the dilution.  Therefore, this approach has been
discussed in the context of the Migdal-Kadanoff
approximation~\cite{BoCo} to justify the scaling Ansatz. Here, we
apply this procedure to the EA in $d=2$, which has been studied
extensively in recent years. These studies have found the value of the
stiffness exponent $y$ to be $y_2=-0.281(2)$~\cite{RSBDJ},
$y_2=0.287(4)$~\cite{HBCMY}, or $y_2=-0.282(2)$~\cite{HY,CBM}. We find
that the result obtained here on diluted lattices, $y_2=-0.281(3)$,
compares well with those earlier results.  This does not add any
accuracy to the value of $y_2$, and we find anew~\cite{BoCo} that
diluted lattices with continuous (Gaussian) bonds are beset with more
complex scaling behavior~\cite{BF} as well as more extensive scaling
corrections. Much more important, though, the present study indicates
the correctness of the reduction approach, hence the validity of the
results obtained for larger dimensions. In particular, the stiffness
exponent does not depend on the dilution, i.e.\ it is universal, even
when $y<0$.

In the next section, we describe the algorithm we applied to reduce
and evaluate large instances of dilute, planar lattices and outline
the ground-state algorithm. In Sec.~\ref{numerics}, we discuss the
results of our numerical studies, followed by some conclusions in
Sec.~\ref{conclusion}.

\section{Algorithms}
\label{algorithms}
In this section we explain the algorithms used to calculate exact
ground states of diluted two-dimensional Ising spin glasses. Our
approach consists of two steps, each previously introduced in
Refs.~\cite{Stiff2} and~\cite{HY}, which we will review in the
following. First, the systems are {\em reduced}, i.e.\ low-connected
spins are iteratively removed, while altering the remaining
interactions such that the ground-state energy is not affected. After
the reduction is finished, the ground state of the remaining system is
calculated exactly using a matching algorithm.

Absent a true glassy state in $d=2$, it is not too surprising that
computationally efficient ground-state algorithms exist
\cite{bieche1980,SG-barahona82b,derigs1991} which exhibit a running
time growing only polynomially with system size. This allows to
measure $y$ with great accuracy. For $d\geq3$, where a true glassy
state exists at low temperatures, no computationally efficient methods
are known to determine ground states exactly. The ground-state
calculation belongs~\cite{Barahona} to the class of {\em NP-hard}
problems~\cite{garey1979}, where all existing algorithms exhibit an
exponentially growing running time with size.  Instead, heuristic
optimization methods \cite{PH-opt-phys2001,PH-opt-phys2004} are used
which typically are believed to approximate ground states for lattices
with up to $n\approx10^3$ spins (or $L\leq 14$ in $d=3$) with some
confidence. Any inaccuracy in the determination of such ground states
gets further aggravated by way of the subtraction leading to $\Delta
E$, suggesting that the scaling regime in $d=3$ extends at most up to
$L\approx10$, and even less in $d>3$.

In light of those difficulties, it might come as a surprise that the
study of the EA on {\it diluted} lattices would possibly improve
matters. After all, dilution eases the constraintness of the spin
configuration, leading to less frustration, and locally to a less
glassy state at low temperatures. Consequently, the length scale
beyond which frustration effects local spin arrangements should be
extended for increasing dilution, leading to persistent scaling
corrections before an asymptotic scaling regime can be obtained at
much larger system sizes. Thus, any gain in obtainable system size
provided by the dilution should only marginally effect any useful
scaling regime. Yet, the numerical results using a $\pm J$
bond-distribution prove otherwiseRefs.~\cite{Stiff1,Stiff2}: While
scaling corrections worsen as expected at too small bond-densities,
they are significantly {\it suppressed} at intermediate densities even
compared to the undiluted case.  The origin of those reduced scaling
corrections at intermediate densities has been investigated in
Ref.~\cite{BoCo}. Additionally, collapsing all data from various bond
fractions $p$ with a scaling Ansatz extends scaling even further.

\subsection{Reduction Method}
\label{reduction}
To exploit the advantages of spin glasses on a bond-diluted lattice,
we can often {\it reduce} the number of relevant degrees of freedom
substantially before a call to an optimization algorithm becomes
necessary. Such a reduction, in particular of low-connected spins,
leads to a smaller, compact remainder graph, bare of trivially
fluctuating variables, which is easier to optimize.  Here, we focus
exclusively on the reduction rules for the energy at $T=0$; a subset
of these also permit the exact determination of the entropy and
overlap~\cite{MKpaper}. These rules apply to general Ising spin glass
Hamiltonians as in Eq.~(\ref{Heq}) with {\it any} bond distribution
$P(J)$, discrete or continuous, on arbitrary sparse graphs.

The reductions effect both spins and bonds, eliminating recursively
all zero-, one-, two-, and three-connected spins. From previous
applications~\cite{Stiff1,Stiff2}, we have supplemented these rules
with one that is not topological but concerns bond values directly,
which is especially effective for Gaussian bond distributions. These
operations eliminate and add terms to the expression for the
Hamiltonian in Eq.~(\ref{Heq}), but leave it form-invariant. Offsets
in the energy along the way are accounted for by a variable $H_o$,
which is {\it exact} for a ground-state configuration.

{\it Rule I:} An isolated spin can be ignored entirely.

{\it Rule II:} A one-connected spin $i$ can be eliminated, since its
state can always be chosen in accordance with its neighboring spin $j$
to satisfy the bond $J_{i,j}$. For its energetically most favorable
state we adjust $H_o:=H_o-|J_{i,j}|$ and eliminate the term
$-J_{i,j}\,x_i\,x_j$ from $H$.

{\it Rule III:} A double bond, $J_{i,j}^{(1)}$ and $J_{i,j}^{(2)}$,
between two vertices $i$ and $j$ can be combined to a single bond by
setting $J_{i,j}= J_{i,j}^{(1)}+J_{i,j}^{(2)}$ or be eliminated
entirely, if the resulting bond vanishes. This operation is very
useful to lower the connectivity of $i$ and $j$ at least by
one.

{\it Rule IV:} For a two-connected spin $i$, rewrite
in Eq.~(\ref{Heq})
\begin{eqnarray}
x_i(J_{i,1}x_1+J_{i,2}x_2)&\leq&\left|J_{i,1}x_1+J_{i,2}x_2\right|\nonumber\\
&=&J_{1,2}x_1x_2+\Delta H,
\label{2coneq}
\end{eqnarray}
where
\begin{eqnarray}
J_{1,2}&=&\frac{1}{2}\left(\left|J_{i,1}+J_{i,2}\right|-\left|J_{i,1}-J_{i,2}\right|\right),\nonumber\\
\Delta H&=&\frac{1}{2}\left(\left|J_{i,1}+J_{i,2}\right|+\left|J_{i,1}-J_{i,2}\right|\right),
\label{h0eq}
\end{eqnarray}
leaving the graph with a new bond $J_{1,2}$ between spin $1$ and $2$,
and acquiring an offset $H_o:=H_o-\Delta H$.

\begin{figure}
\vskip 1.2in \includegraphics{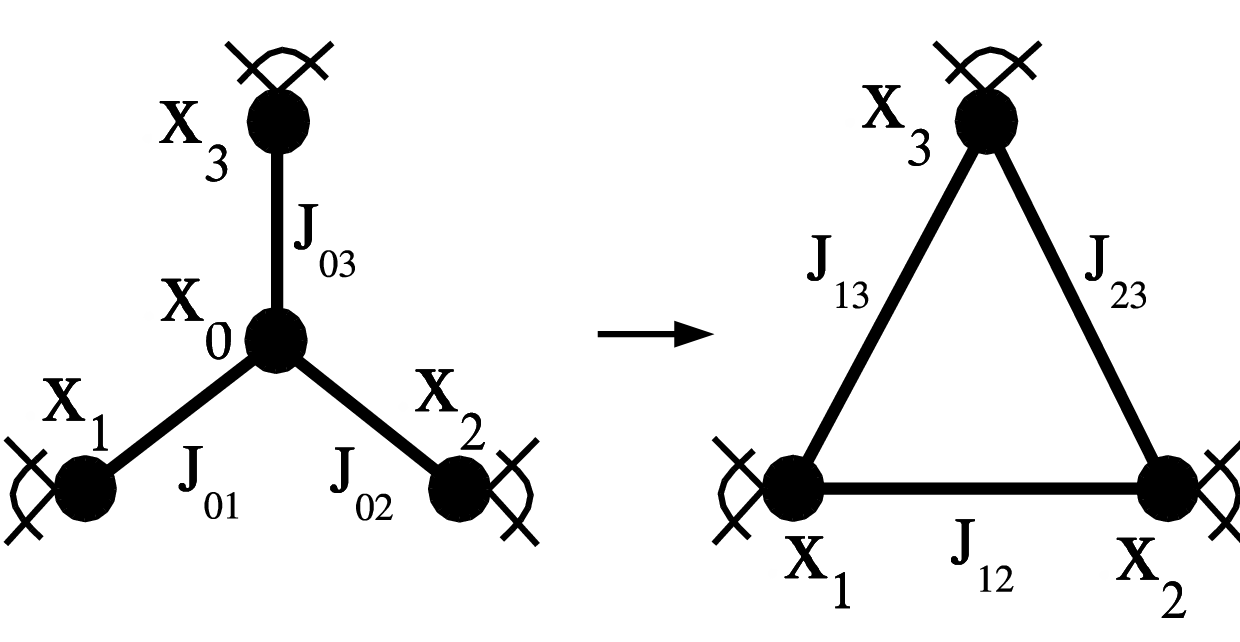}
\caption{``Star-triangle'' relation to reduce a three-connected spin
$x_0$. The new bonds on the right are obtained in
Eq.~(\protect\ref{3coneq}).  }
\label{startri}
\end{figure}

{\it Rule V:} A three-connected spin $i$ can be reduced via a
``star-triangle'' relation, as depicted in Fig.~\ref{startri}:
\begin{eqnarray}
&&J_{i,1}\,x_i\,x_1+J_{i,2}\,x_i\,x_2+J_{i,3}\,x_i\,x_3\nonumber\\
&\leq&\left|J_{i,1}x_1+J_{i,2}x_2+J_{i,3}x_3\right|\\
&=&J_{1,2}\,x_1\,x_2+J_{1,3}\,x_1\,x_3+J_{2,3}\,x_2\,x_3+\Delta H,\nonumber
\label{3coneq}
\end{eqnarray}
where
\begin{eqnarray}
&J_{1,2}=-A-B+C+D,\quad J_{1,3}=A-B+C-D,&\nonumber\\
&J_{2,3}=-A+B+C-D,\quad \Delta H=A+B+C+D,&\nonumber\\
&A=\frac{1}{4}\left|J_{i,1}-J_{i,2}+J_{i,3}\right|, \quad
 B=\frac{1}{4}\left|J_{i,1}-J_{i,2}-J_{i,3}\right|,&\nonumber\\
&C=\frac{1}{4}\left|J_{i,1}+J_{i,2}+J_{i,3}\right|,\quad
 D=\frac{1}{4}\left|J_{i,1}+J_{i,2}-J_{i,3}\right|.&\nonumber
\label{h1eq}
\end{eqnarray}

\begin{figure}
\vskip 1.7truein \includegraphics{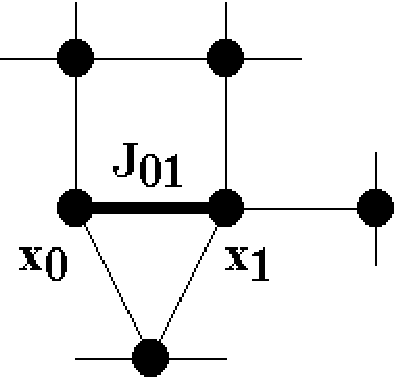} \includegraphics{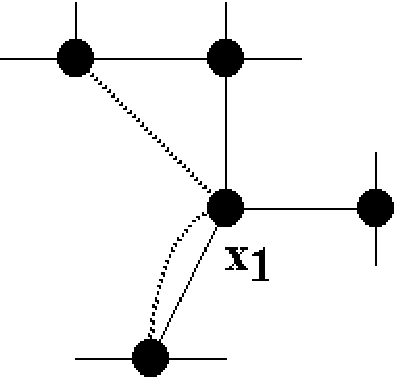}
\caption{Illustration of {\it Rule VI} for ``strong'' bonds. Left, the
  local topology of a graph is shown for two spins, $x_0$ and $x_1$,
  connected by a bond $J_{0,1}$ (thick line). If $J_{0,1}>0$
  (resp. $J_{0,1}<0$) satisfies Eq.~(\protect\ref{bondeq}), $x_0$ and
  $x_1$ must align (resp. anti-align) in the ground state and $x_0$
  can be removed. Right, the remainder graph is show after the
  removal. The other bonds emanating from $x_0$ (dashed lines) are now
  directly connected to $x_1$ (potentially with a sign change, if
  $J_{0,1}<0$). This procedure may lead to a double bond ({\it Rule
  III}), if $x_1$ was already connected to a neighbor of $x_0$
  before.}
\label{superbond}
\end{figure}

{\it Rule VI:} A spin $i$ (of any connectivity) for which the absolute
weight $|J_{i,j'}|$ of one bond to a spin $j'$ is larger than the
absolute sum of all its other bond-weights to neighboring spins
$j\not=j'$, i. e.
\begin{eqnarray}
|J_{i,j'}|>\sum_{j\not=j'}|J_{i,j}|,
\label{bondeq}
\end{eqnarray}
bond $J_{i,j'}$ {\it must} be satisfied in any ground state. Then,
spin $i$ is determined in the ground state by spin $j'$ and it as well
as the bond $J_{i,j'}$ can be eliminated accordingly, as depicted in
Fig.~\ref{superbond}. Here, we obtain $H_0:=H_0-|J_{i,j'}|$. All other
bonds connected to $i$ are simply reconnected with $j'$, but with
reversed sign, if $J_{i,j'}<0$.

This procedure is costly, and hence best applied after the other rules
are exhausted. But it can be highly effective for very widely
distributed bonds. In particular, since neighboring spins may reduce
in connectivity and become susceptible to the previous rules again,
further reductions may ensue, see Fig.~\ref{superbond}.

The bounds in Eqs.~(\ref{2coneq}-\ref{3coneq}) become {\it exact} when
the remaining graph takes on its ground state.  Reducing even
higher-connected spins would lead to new (hyper-)bonds between more
than two spins, unlike Eq.~(\ref{Heq}). While such a reduction is
possible and would eventually result in the complete evaluation of any
lattice ground state, it would lead along the way to an exponential
proliferation in the number of (hyper-)bonds in the system. This fact
is a reflection of the combinatorial complexity of the glassy state,
which will be explored in Ref.~\cite{BoDu}.

After a recursive application of these rules, the original lattice
graph is either completely reduced (which is almost always the case
below or near $p_c$), in which case $H_o$ provides the exact ground
state energy already, or we are left with a highly reduced, compact
graph in which no spin has less than four connections. We obtain the
ground state of the reduced graph with an exact matching algorithm as
used in Ref.~\cite{HY}, which together with $H_o$ provides the
ground-state energy of the original diluted lattice instance.

\begin{figure}[htb] 
\includegraphics[angle=0,width=0.95\columnwidth]{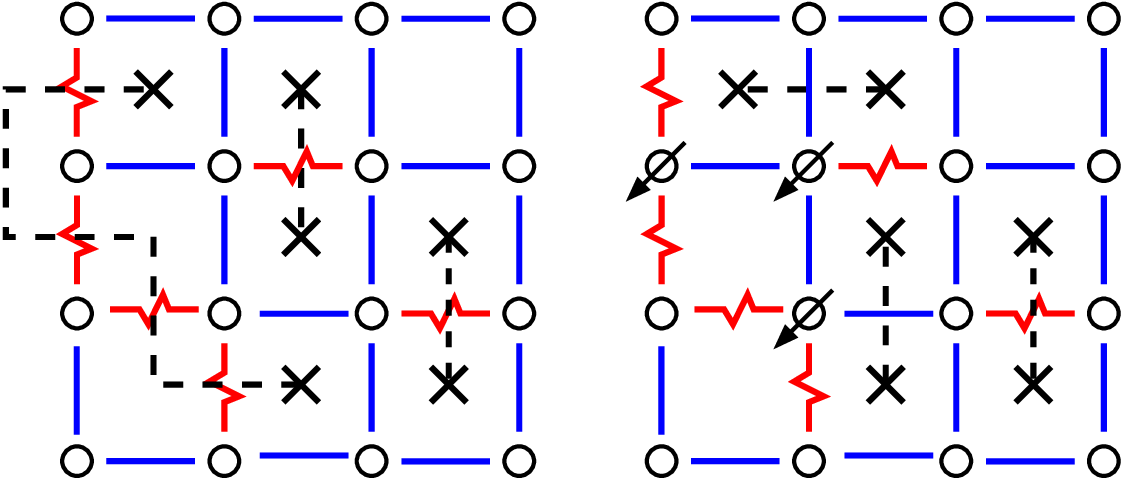}
\caption{2d spin glass with all spins up (left, up spins not
  shown). Straight lines are ferromagnetic, jagged lines are
  anti-ferromagnetic bonds. The dotted lines connect frustrated
  plaquettes (crosses).  The bonds crossed by the dotted lines are
  unsatisfied. In the right part the ground state with three spins
  pointing down (all other up) is shown, corresponding to a minimum
  number of unsatisfied bonds.}
\label{fig:matching}
\end{figure}

\subsection{Matching}
\label{matching}
Let us now explain just the basic idea of the matching algorithm, for
the details, see Refs.~\cite{bieche1980,SG-barahona82b,derigs1991}.
The method works for spin glasses which are planar graphs; this is the
reason, why we apply periodic boundary conditions only in one
direction. In the left part of Fig.~\ref{fig:matching} a small 2d
system with open boundary conditions is shown. All spins are assumed
to be ``up'', hence all anti-ferromagnetic bonds are not satisfied. If
one draws a dotted line perpendicular to all unsatisfied bonds, one
ends up with the situation shown in the figure: all dotted lines start
or end at frustrated plaquettes and each frustrated plaquette is
connected to exactly one other frustrated plaquette. Each pair of
plaquettes is then said to be {\em matched}. Now, one can consider the
frustrated plaquettes as the vertices and all possible pairs of
connections as the edges of a (dual) graph.  The dotted lines are
selected from the edges connecting the vertices and called a {\em
perfect} matching, since {\em all} plaquettes are matched.  One can
assign weights to the edges in the dual graph, the weights are equal
to the sum of the absolute values of the bonds crossed by the dotted
lines. The weight $\Lambda$ of the matching is defined as the sum of
the weights of the edges contained in the matching. As we have seen,
$\Lambda$ measures the broken bonds, hence, the energy of the
configuration is given by $E=-\sum_{\langle i,j\rangle}
|J_{ij}|+2\Lambda$. Note that this holds for {\em any} configuration
of the spins, since a corresponding matching always exists.  Although
Fig.~\ref{fig:matching} only shows a square lattice, a matching is
always possible for {\it any} planar graph, such as the reduced,
dilute lattices discussed here.

Obtaining a ground state means minimizing the total weight of the
broken bonds (see right panel of Fig.~\ref{fig:matching}), so one is
looking for a {\em minimum-weight perfect matching}. This problem is
solvable in polynomial time.  The algorithms for minimum-weight
perfect matchings~\cite{MATCH-cook,MATCH-korte2000} are among the most
complicated algorithms for polynomial problems.  Fortunately the LEDA
library offers a very efficient implementation~\cite{PRA-leda1999},
which we have applied here.

\section{Scaling Ansatz}
\label{scaling}
Clearly, there exists a lowest bond fraction $p^*$, below which a
glassy state is not possible. In particular, the lattice must exceed
the bond-percolation threshold $p_c$ to exhibit any long-range
correlated behavior and $p^*\geq p_c$ must hold.  It is expected that
$p^*=p_c$ for any continuous distribution, while for discrete
distributions $p^*$ may be minutely larger than
$p_c$~\cite{BF,MKpaper,Stiff1}. Accordingly, for two-dimensional
bond-diluted lattices and a Gaussian bond distribution used here, we
expect $p^*=p_c=1/2$~\cite{A+S}.  Similarly, it is expected that the
correlation length near the critical point scales as
\begin{equation}
\label{eq:corr-length}
\xi(p)\sim(p-p^*)^{-\nu^*}\quad\,
\end{equation}
with $\nu^*=\nu=4/3$, well-known from 2-dimensional percolation.

The introduction of the bond density $p$ as new parameter permits a
finite-size-scaling Ansatz in the limit $p\to p^*$. Combining the data
for all $L$ {\it and} $p$ leads to a new variable $x=L/\xi(p)$, which
has the chance of exhibiting scaling over a wider regime than for $L$
alone. As we have argued in Ref.~\cite{BoCo}, this Ansatz should take
the form
\begin{eqnarray}
\sigma(\Delta E)_{L,p}\sim \xi(p)^{y_P}~x^y~f(x),
\label{extendedyeq}
\end{eqnarray}
as suggested by Refs.~\cite{BBF,BF}.  In principle, the Ansatz
requires $L\gg1$ and $\xi(p)\gg1$. The scaling function $f$ was chosen
as to approach a constant for $L\gg\xi(p)$.

Note that one basic assumption used here is that the exponent $y$ does
not depend on the bond density $p$. Since a percolating cluster {\it
sufficiently above} the percolation transition is effectively compact
for large $L$, one may argue that asymptotic scaling properties of the
spin glass should be uneffected by $p$. Reproducing the scaling of the
undiluted lattice with the Ansatz in Eq.~(\ref{extendedyeq}) with some
accuracy would add support to this argument.

To obtain the exponent $y_P$ in Eq.~(\ref{extendedyeq}) directly, one
considers the limit $L\sim\xi(p)\to\infty$, i.~e. $x\sim1$. Then, one
can show that at $p=p_c$~\cite{BBF,BoCo}
\begin{eqnarray}%
\sigma(\Delta E)_{L,p_c}\sim L^{y_P}.%
\label{yPeq}%
\end{eqnarray}%
We observe that due to the fractal nature of the percolating cluster
at $p^*=p_c$, no long-range order can be sustained and defects possess
a vanishing interface. Thus, $y_P\leq0$ and the defect energy
vanishes.  In contrast, $y_P=0$ for discrete bond distributions such
as $\pm J$, because $p^*>p_c$ and the interface energy $\sigma(\Delta
E)_{L,p^*}$ becomes $L$-independent.  As it turns out, with $y_P=0$
the scaling collapse greatly simplifies for $\pm J$ bonds, leaving
continuous bonds with one extra exponent to account for. Hence, using
Gaussian bonds, the accuracy obtained for the desired stiffness
exponent $y$ diminishes, as the study in Ref.~\cite{BoCo} shows.

In case of a finite-temperature glass transition with divergent energy
scales ($y>0$), universality provides us with the choice of the more
convenient distribution, $\pm J$, to compute the stiffness
exponent~\cite{Stiff1,Stiff2}. Apparently, this universality brakes
down below the lower critical dimension,
$d<d_l\approx2.5$~\cite{AMMP,lowd}, and a nontrivial value for $y$ is
only obtained for continuous bonds~\cite{HY}. Thus, we have to take
the exponent $y_P$ in Eq.~(\ref{extendedyeq}) into account.

A scaling collapse is further complicated by the fact that the
asymptotic regime of interest for the determination of $y$, namely
$x\gg1$ or $L\gg\xi(p)$, is hard to access for $p\to p^*$. Most data
that reaches asymptotic scaling, i.~e. $x\gg1$, is typically obtained
instead at intermediate values of $p$, sufficiently above $p^*$ to
reach system sizes with $L\gg\xi(p)$ but small enough to exploit the
reduction rules from Sec.~\ref{reduction}.  As the analysis in
Ref.~\cite{BoCo} suggest, in that regime the correlation length
Eq.~(\ref{eq:corr-length}) may be too small to justify
Eq.~(\ref{extendedyeq}). Furthermore, it is clear that
Eq.~(\ref{eq:corr-length}) is valid only close to $p_c$, leading
possibly to wrong estimations for critical exponents obtained from
finite-size scaling~\cite{carter2003}. Also the finite-size
corrections for the correlation length itself play probably an
important role, which can bee seen from previous
studies~\cite{eschenbach1981} of two-dimensional percolation, where a
significant change of the effective exponent $\nu$ was observed when
changing the system size from $L=2$ to $L=10000$. Similarly, unknown
scaling corrections missing in the form of Eq.~(\ref{extendedyeq}) are
likely to arise. Yet, experience shows that a focus on data with
$L\gg\xi(p)$ for any $\xi(p)$ at least provides a satisfactory
collapse in the $x\gg1$ regime with an accurate prediction for the
stiffness exponent $y$. There, unlike $y$, the exponents $\nu^*$ and
$y_P$, and the scaling function $f(x)$, which are more closely
associate with the scaling window near $p^*$, will {\it not} be
accurately represented. In any fit of the data, their values are
likely distorted to absorb the effect of unknown scaling corrections.

For our data analysis, we will therefore apply cuts to eliminate data
outside of $x\gg1$, and fit the remaining data to the form
\begin{eqnarray}
\frac{\sigma(\Delta E)_{L,p}}{\xi(p)^{y_P}}\sim
\left[\frac{L}{\xi(p)}\right]^y~f(\infty),
\label{fiteq}
\end{eqnarray}
fixing $\xi(p)=(p-p^*)^{-\nu^*}$, with $f(\infty)$, $p^*$, $\nu^*$,
$y_P$, and $y$ as fitting parameters.  Note that in this limit,
$\nu^*$ and $y_P$ are {\it not} independent. Hence, we choose to fix
$\nu^*=\nu=4/3$.

\section{Numerical Experiments}
\label{numerics}
In our numerical simulation, we have generated a large number of
instances of symmetric Gaussian bond disorder on square lattices with
open boundaries vertically, and periodic boundary conditions
horizontally, as described in Ref.~\cite{HY}. But these instances are
bond-diluted with a bond fraction of $p\geq p^*=p_c=1/2$. On this
bond-diluted spin glass, the reduction algorithm from
Sec.~\ref{reduction} is applied to recursively remove as many spin
variables as possible while exactly accounting for their contribution
to the ground-state energy. Since the original lattice was a planar
graph, the reduction method preserves this property for the remainder
graph. Hence, the matching algorithm discussed in Sec.~\ref{matching} can
be applied to the remainder graphs here to determine their exact
ground states in polynomial time.  In this manner, we study the defect
energy $\sigma(\Delta E)$ both as a function of size $L$ {\it and}
bond density $p$. We consider systems of sizes from maximally $L=150$
at $p=1$ to up to maximally $L=1000$ at $p=0.52$. At each pair of $L$
and $p$ we average typically well over about $10^4$ instances.

\begin{figure}
\vskip 2in \includegraphics{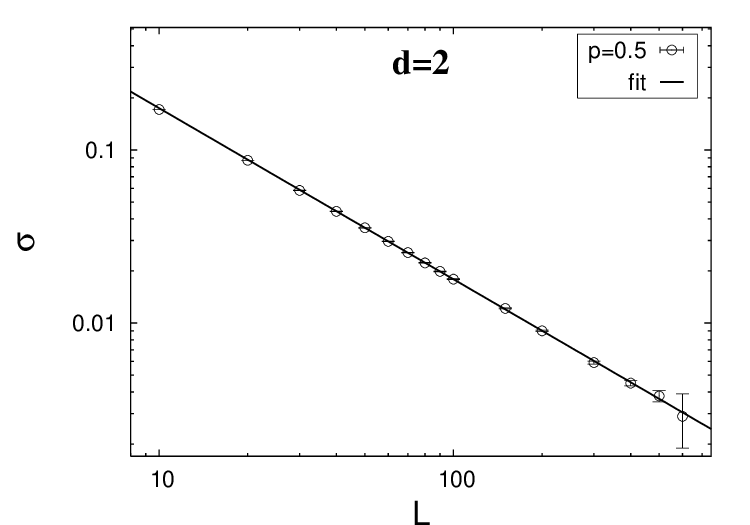}
\caption{Logarithmic plot of $\sigma_L$ at $p=p^*=1/2$. An asymptotic
  power-law regime corresponding to Eq.~(\protect\ref{yPeq}),
  different from those for $p>p^*$, is reached quickly, and an
  asymptotic fit extrapolates to $y_P=-0.98(2)$.  }
\label{yPplot}
\end{figure}

Before we proceed to collapsing the data, it is instructive first to
determine the exponent $y_P$ directly according to
Eq.~(\ref{yPeq}). Clearly, at $p=p^*$, diluted lattices are almost all
entirely reducible with the rules given in Sec.~\ref{reduction}, and
rarely is any subsequent optimization necessary. Hence, $L$ is limited
only by the cost of reduction itself, memory space, and
statistics. The data for the defect energy at $p^*=1/2$ is plotted in
Fig.~\ref{yPplot}. The asymptotic fit yields about
$y_P=-0.98(2)$. This value seems to suggest that $y_P=-1$, which would
imply that the spin glass on a percolating cluster in two dimensions
is essentially one-dimensional (where $y=-1$).  Concluding that
$y_P=-1$ is exact may be misguided: Ref.~\cite{BBF} obtained
$y_P\approx-0.99$ on the basis of a scaling argument involving the
numerical solution of an integral.

\begin{figure}
\vskip 2in \includegraphics{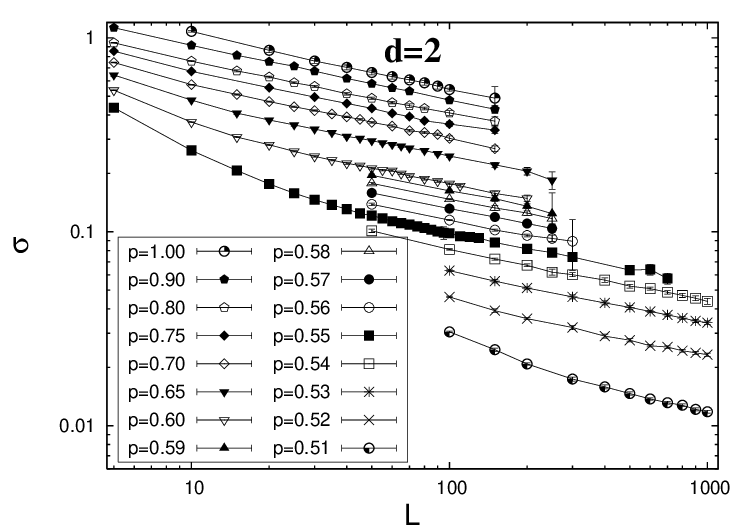}
\caption{Plot on a logarithmic scale of the stiffness $\sigma$ as a
function of system size $L$.  The data is grouped into sets (connected
by lines) parameterized by the bond density $p$. Most sets show a
distinct scaling regime as indicated by Eq.~(\protect\ref{yeq}) for
sufficiently large $L$. In particular, the data for $p=0.51$ appears
to never reach scaling and will be disregarded entirely.  }
\label{rawdefectplot}
\end{figure}

\begin{figure}
\vskip 2in \includegraphics{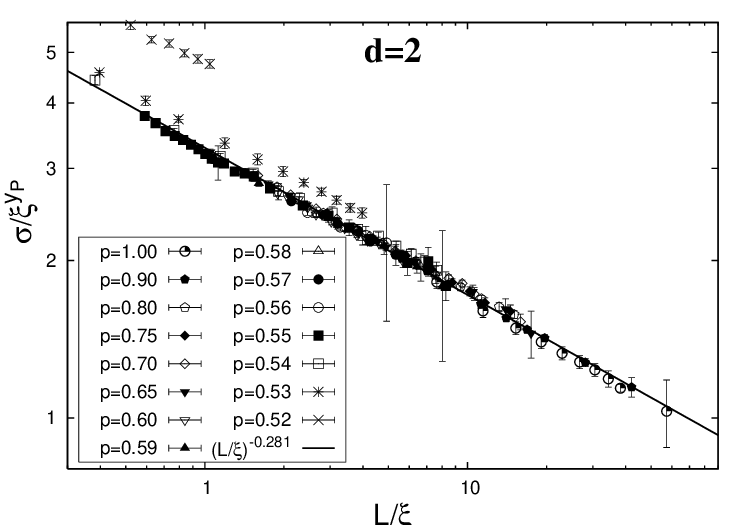}
\caption{Scaling plot of the data from
Fig.~\protect\ref{rawdefectplot} for $\sigma$, fitted to
Eq.~(\protect\ref{fiteq}) as a function of the scaling variable
$x=L/\xi(p)$ where $\xi(p)=(p-p^*)^{-\nu^*}$. Data below the scaling
regime in each set (i.~e. below a certain value of $L$ for each $p$)
from Figs.~\protect\ref{rawdefectplot} was cut. The continuous line
represent the power-law $[L/\xi(p)]^{-0.281}$, using the fitted value
for $y$. }
\label{scaldefectplot}
\end{figure}

To obtain an optimal scaling collapse of the data in accordance with
the discussion in Sec.~\ref{scaling}, we focus on the data in the
asymptotic scaling regime for each set. To this end, we chose for each
data set a lower cut in $L$ by inspection of the data in
Fig.~\ref{rawdefectplot}. All data points below the cut for each $p$
are discarded, all data above are kept. Then the remaining data for
all $L$ and $p$ are fitted to the four-parameter scaling form in
Eq.~(\ref{fiteq}). The resulting collapse is displayed in
Fig.~\ref{scaldefectplot}.

Initiating all of the parameters at near-optimal values, a
least-square fit incorporating all the data shown in the collapse
converges. The fitted values are $y=-0.281(3)$, $y_P=-0.77(5)$,
$p^*=0.514(5)$, and $f(\infty)\approx3.3$.  Errors in this fit are
estimates based the sensitivity on varying each parameter and have to
be judged cautiously. It should be noted how essential the inclusion
of the parameter $y_P$ was for the collapse, even obtaining a fitted
value not too far from its actual value determined in
Fig.~\ref{yPplot}.  The discrepancy between the fitted value
$y_P=-0.77$ and the accurate value $y_P=-0.99$ is due to the unknown
scaling for $\xi(p)$ away from $p_c$ and the scaling-corrections for
the approach Eq.~(\ref{extendedyeq}), as discussed above.

\section{Conclusion}
\label{conclusion}
We have used a scaling Ansatz proposed in Refs.~\cite{BBF,BF} in
conjunction with the spin reduction scheme of
Refs.~\cite{Stiff1,Stiff2,BoCo} and exact ground-state calculations to
study the defect energy at $T=0$ for bond-diluted lattices in two
dimensions.  The results for the stiffness exponent $y$ scale over 2
decades and are consistent with previous studies, validating the basic
Ansatz. Yet, the obtained value $y=-0.281(3)$ is only of comparable
accuracy to those studies, and the data collapse provides less of an
advance than bond-diluted lattices did for higher-dimensional
lattices. For one, in two dimensions there is no finite-temperature
glass transition ($y<0$) and conventional studies at full connectivity
are successful at reaching large lattice sizes already, avoiding the
additional uncertainties of a multi-parameter fit. In this scenario,
Ref.~\cite{BoCo} argued that such a collapse of data may provide
diminishing returns for the computational effort. Similarly, it was
observed there that a continuous bond distribution, considering the
smaller size of elementary excitations under bond-reversal, leads to
larger scaling corrections in $L$.  Those scaling corrections are
enhanced further by open boundaries, which have been observed
previously to result in only weakly decreasing
corrections~\cite{HY,Middle,HM}.

Our results indicate the validity of the recently proposed reduction
scheme to determine the stiffness exponent $y$.  Note that the fact
that reduction works well in two-dimensions, where $T_c=0$ holds, does
not imply definitely that it should work for $d>2$. Nevertheless,
since the overall behavior in $d=2$ and higher dimensions is similar,
it is highly probable that the reduction scheme is applicable also for
higher dimensions~\cite{Stiff1,Stiff2}, where no exact ground-state
algorithms are available.

\section*{Acknowledgments}
SB is supported by grant 0312510 from the Division of Materials
Research at the National Science Foundation and a grant from the Emory
University Research Council. AKH has obtained financial support from
the {\em VolkswagenStiftung} (Germany) within the program
``Nachwuchsgruppen an Universit\"aten'' and from the European
Community via the DYGLAGEMEM program.

\end{document}